\documentclass{article}
\usepackage[T1]{fontenc}
\usepackage[latin1]{inputenc}
\usepackage{a4wide}
\makeatletter

\providecommand{\LyX}{L\kern-.1667em\lower.25em\hbox{Y}\kern-.125emX\@}

 \newcommand{\lyxaddress}[1]{
   \par {\raggedright #1 
   \vspace{1.4em}
   \noindent\par}
 }

\usepackage[T1]{fontenc}
\usepackage[latin1]{inputenc}
\usepackage{a4wide}


\begin{document}

\title{Entanglement vs. Noncommutativity in Teleportation}

\author{Sibasish Ghosh\( ^{\star } \)\thanks{
res9603@isical.ac.in 
}, Guruprasad Kar\( ^{\star } \)\thanks{
gkar@isical.ac.in 
}, Anirban Roy\( ^{\star } \)\thanks{
res9708@isical.ac.in 
} and Ujjwal Sen\( ^{\natural } \)\thanks{
ujjwalsen@yahoo.co.in
}}

\maketitle

\lyxaddress{\( ^{\star } \)Physics and Applied Mathematics Unit, Indian Statistical Institute,
203 B.T. Road, Kolkata 700035, India}

\lyxaddress{\( ^{\natural } \)Department of Physics, Bose Institute, 93/1 A.P.C. Road,
Kolkata 700009, India}

\begin{abstract}
We provide an alternative simple proof of the necessity of entanglement in quantum
teleportation by using the no-disentanglement theorem. We show that this is
true even when the state to be teleported is known to be among two noncommuting
qubits. We further show that to teleport any set of commuting qubits, it is
sufficient to have a classically correlated channel. Using this result we provide
a simple proof of the fact that any set of bipartite entangled states can be
exactly disentangled if the single particle density matrices of any one party
commute. 
\end{abstract}
The idea of quantum teleportation is to send an unknown state to a distant party
without actually sending the particle itself using only local operations and
classical communication (LOCC) between them. A protocol for this scheme was
proposed by Bennett \textit{et al.} \cite{1}, where a maximally entangled state
is required as channel state between the two parties. In teleporting a state
from one party to another where only local operations and classical communication
is allowed between the two parties, the question of necessity of entanglement
of the channel state is a fundamental one. This issue has been discussed in
a sketchy way in \cite{1} as well as in \cite{2}. It is in Ref. \cite{3,4}
where this issue has been discussed in a somewhat detailed manner. In this letter
we shall discuss this issue of necessity of entanglement of the channel state
for \textit{exact} quantum teleportation of a given set of states of a single
qubit. In this direction we provide an alternative simple reasoning to show
that for universal teleportation, one necessarily requires an entangled channel
state. Next we show that entanglement of the channel is necessary even to teleport
any set of noncommuting qubits. These proofs are \textit{independent} of any
teleportation protocol. We then provide a protocol by which any set of commuting
states can be teleported through a classically correlated channel. This allows
us to give a simple proof of the fact that the entangled states of two qubits,
whose reduced density matrices of one party commute, can be disentangled exactly. 

We first provide a simple reasoning as to why entanglement of the channel state
is necessary for exactly teleporting an unknown qubit. Consider a separable
channel state between two distant parties, Alice and Bob. Suppose that it is
possible to teleport (exactly) an arbitrary qubit (call it qubit 1) from Alice
to Bob through this channel state. Now this qubit may be one part of a two qubit
entangled state \( \rho _{12} \) at Alice's side. As Alice and Bob do not possess
any shared entanglement before implementation of the teleportation protocol,
they would not share any after it. Therefore after the teleportation protocol,
the initial entangled state \( \rho _{12} \) would get transformed into a separable
one with reduced density matrices remaining intact \cite{5}. But this would
make universal disentanglement possible contradicting the No-Disentanglement
theorem \cite{6}. So to teleport exactly a universal set of qubits, entanglement
of the channel state is necessary \cite{9}.

We now show that entanglement of the channel is necessary even for exactly teleporting
any set of two nonorthogonal states \cite{11}.

Consider the following set of bipartite normalized states, \[
{{\mathcal{F}}}=\{|0\alpha \rangle ,|1\beta \rangle ,\frac{1}{\sqrt{2}}(|0\alpha \rangle +|1\beta \rangle )\},\]
 where \( |0\rangle  \), \( |1\rangle  \) are orthogonal states and \( |\alpha \rangle  \),
\( |\beta \rangle  \) are nonorthogonal states. We first show below that the
set \( {{\mathcal{F}}} \) can not be exactly disentangled into separable states
\cite{12}, in a fashion similar to that in Mor \cite{7}. If possible, let
there exist a unitary operator \( U \), acting on these states together with
a fixed ancilla state \( |A\rangle  \), realizing exact disentanglement of
these states. So we must have \[
U(|0\alpha A\rangle )=|0\alpha A_{0}\rangle ,\]
 \[
U(|1\beta A\rangle )=|1\beta A_{1}\rangle ,\]
 \[
U(\frac{1}{\sqrt{2}}(|0\alpha A\rangle +|1\beta A\rangle ))=\frac{1}{\sqrt{2}}(|0\alpha A_{0}\rangle +|1\beta A_{1}\rangle ),\]
 where the ancilla states should satisfy the relation \( \langle A_{0}|A_{1}\rangle =1 \)
(to keep the reduced density matrices intact). This will not change the entanglement
of the state \( \frac{1}{\sqrt{2}}(|0\alpha \rangle +|1\beta \rangle ) \) at
all. Hence the set \( {{\mathcal{F}}} \) cannot be exactly disentangled. (This
result also shows that universal disentanglement is not possible.) Interestingly,
Mor \cite{7} proved the same result using a set of four states, and conjectured
that this result can be proved with fewer (\textit{i.e.}, less than four) states.

Consider now a separable channel state between two distant parties, Alice and
Bob. Assume that the set \( \{|\alpha \rangle  \), \( |\beta \rangle \} \)
of two nonorthogonal states can be teleported exactly from Alice to Bob through
this channel state. Hence any mixture of \( P[|\alpha \rangle ] \) and \( P[|\beta \rangle ] \)
can also be teleported through this channel state, as the channel keeps no imprint
of the states \( |\alpha \rangle  \) or \( |\beta \rangle  \) after their
teleportation. Therefore Alice would be able to (exactly) teleport to Bob, the
state of the second particle of an arbitrary two-particle state chosen from
the set \( {{\mathcal{F}}}=\{|0\alpha \rangle ,|1\beta \rangle ,(1/\sqrt{2})(|0\alpha \rangle +|1\beta \rangle \} \).
Since the channel state is separable, the set \( {{\mathcal{F}}} \) would get
exactly disentangled. But this is impossible, as was shown above. So we conclude
that \textit{entanglement of the channel is necessary for exact teleportation
of a state known to be among two given nonorthogonal states}.

Next we show that teleportation any two noncommuting qubits also requires an
entangled channel. We shall require the following lemma.

\textit{Lemma}: The set \( {{\mathcal{S}}}=\{{\rho }_{AB}^{1},{\rho }_{AB}^{2}\} \)
of two \( 2\otimes 2 \) states, where at least one of them is entangled, cannot
be exactly disentangled by applying any physical operation on the side \( B \),
if the reduced density matrices on the side \( B \) do not commute.

\textit{Proof}: As the reduced density matrices on the side \( B \) do not
commute, there exist two nonorthogonal pure qubits \( |\psi \rangle  \) and
\( |\phi \rangle  \), such that \( tr_{A}({\rho }_{AB}^{j})={\lambda }_{j}P[|\psi \rangle _{B}]+(1-{\lambda }_{j})P[|\phi \rangle _{B}] \)
\( (j=1,2) \), where \( {\lambda }_{1}\ne {\lambda }_{2} \), and at least
one of them is different from both 0 and 1. If possible, let \( U_{BM} \) be
an unitary operator acting on party \( B \) and an ancilla \( M \), attached
with \( B \), realizing the disentangling process. Then after disentanglement,
the joint state of the two parties \( A \) and \( B \) becomes \( {{\rho }^{\prime }}_{AB}^{j}=tr_{M}[I_{A}\otimes U_{BM}{\rho }_{AB}^{j}\otimes P[|M\rangle ]I_{A}\otimes {U}_{BM}^{\dag }] \)
\( (j=1,2) \), \( |M\rangle  \) being the initial state of the ancilla. As
we demand exact disentanglement, we must have \begin{equation}
\label{dis1}
tr_{B}[{\rho }_{AB}^{j}]=tr_{B}[{{\rho }^{\prime }}_{AB}^{j}],
\end{equation}
\begin{equation}
\label{dis2}
tr_{A}[{\rho }_{AB}^{j}]=tr_{A}[{{\rho }^{\prime }}_{AB}^{j}].
\end{equation}
 Eq. (\ref{dis1}) holds trivially, as nothing has been done on party A. Eq.
(\ref{dis2}) gives \[
{\lambda }_{j}P[|\psi \rangle ]+(1-{\lambda }_{j})P[|\phi \rangle ]={\lambda }_{j}tr_{M}[P[U_{BM}(|\psi \rangle \otimes |M\rangle )]]\]
\begin{equation}
\label{lam1}
+(1-{\lambda }_{j})tr_{M}[P[U_{BM}(|\phi \rangle \otimes |M\rangle )]],
\end{equation}
 for \( j=1,2 \). Eq. (\ref{lam1}) will be satisfied if and only if \begin{equation}
\label{lam2}
\left. \begin{array}{lcr}
U_{BM}(|\psi \rangle \otimes |M\rangle ) & = & |\psi \rangle \otimes |M_{0}\rangle ,\\
U_{BM}(|\phi \rangle \otimes |M\rangle ) & = & |\phi \rangle \otimes |M_{1}\rangle .
\end{array}\right\} 
\end{equation}
 Unitarity demands that the ancilla states \( |M_{0}\rangle  \) and \( |M_{1}\rangle  \)
should be identical. Hence none of the states in the set \( {{\mathcal{S}}} \)
will be changed (except a possible change in the identification of the particles)
by this disentangling process. Thus the set \( {\mathcal{S}} \) cannot be exactly
disentangled by applying a disentangling operation on \( B \)'s side.\( \diamondsuit  \)

We now show that \textit{to teleport any set of noncommuting qubits, entanglement
of the channel is necessary}.

Suppose that it is possible to teleport a state chosen at random from the set
\( \{\rho _{1},\rho _{2}\} \) of two noncommuting qubits through an unentangled
channel. If both of the states \( \rho _{1} \), \( \rho _{2} \) are pure,
it has been shown above that they cannot be teleported exactly through an unentangled
channel. So we assume here that at least one of \( \rho _{1} \), \( \rho _{2} \)
is a nonpure state. Since \( \rho _{1} \) and \( \rho _{2} \) are noncommuting,
there uniquely exist two nonorthogonal states \( |\psi \rangle  \) and \( |\phi \rangle  \)
such that \[
\rho _{j}={\lambda }_{j}P[|\psi \rangle ]+(1-{\lambda }_{j})P[|\phi \rangle ]\: (j=1,2),\]
 where \( 0\leq \lambda _{j}\leq 1 \), \( \lambda _{1}\ne \lambda _{2} \),
and at least one of the \( \lambda _{j} \)'s is different from both 0 and 1.
Let us choose two \( 2\otimes 2 \) states \( {\rho }_{AB}^{1} \) and \( {\rho }_{AB}^{2} \)
(at least one of which is entangled), where \( tr_{A}[{\rho }_{AB}^{j}]={\rho }_{j} \),
for \( j=1,2 \). Then the set \( \{{\rho }_{AB}^{1},{\rho }_{AB}^{2}\} \)
can be disentangled exactly by telporting the states of \( B \)'s side through
the unentangled channel. This has been shown to be impossible. So exact teleportation
of any set of noncommuting qubits requires entanglement of the channel.

The obvious next question is whether entanglement of the channel is necessary
even to teleport a set of commuting states. We know that for teleportation of
two orthogonal states, no correlation (quantum or classical) of the channel
is required - a phone call is sufficient. Here we show that for teleportation
of any set of commuting states, a classically correlated channel state is sufficient.

Suppose that Alice has to send any one of the states from the \emph{largest}
set of commuting qubits \( \{wP[|0\rangle _{A_{1}}]+(1-w)P[|1\rangle _{A_{1}}] \)
: \( 0\leq w\leq 1\} \) to Bob (\( \{|0\rangle ,|1\rangle \} \) is a known
orthonormal basis) and Alice and Bob share the \emph{separable} channel state
\( \frac{1}{2}P[|00\rangle _{A_{2}B}]+\frac{1}{2}P[|11\rangle _{A_{2}B}] \)
between them. The three particle state is \[
\frac{wP}{2}[|000\rangle _{A_{1}A_{2}B}]+\frac{(1-w)}{2}P[|111\rangle _{A_{1}A_{2}B}]\]
 \[
+\frac{w}{2}P[|011\rangle _{A_{1}A_{2}B}]+\frac{(1-w)}{2}P[|100\rangle _{A_{1}A_{2}B}].\]
Alice applies a discriminating measurement between the subspaces associated
with the following two-dimensional projectors: \( P_{1}=P[|00\rangle _{A_{1}A_{2}}]+P[|11\rangle _{A_{1}A_{2}}] \)
and \( P_{2}=P[|01\rangle _{A_{1}A_{2}}]+P[|10\rangle _{A_{1}A_{2}}] \). If
\( P_{1} \) clicks, the state of the whole system system becomes \( wP[|000\rangle _{A_{1}A_{2}B}]+(1-w)P[|111\rangle _{A_{1}A_{2}B}] \)
so that Bob's state is \( wP[|0\rangle _{B}]+(1-w)P[|1\rangle _{B} \). Alice
just rings up Bob to tell him the result of her measurement. But if \( P_{2} \)
clicks, Alice informs this to Bob and he has to apply the unitary operator that
converts \( |0\rangle \rightarrow |1\rangle  \) and \( |1\rangle \rightarrow |0\rangle  \)
(\emph{i.e.} \( \sigma _{x} \)) on his particle. In this case the three-qubit
state transforms to \( wP[|010\rangle _{A_{1}A_{2}B}]+(1-w)P[|101\rangle _{A_{1}A_{2}B}] \).
Tracing out \( A_{1} \) and \( A_{2} \), Bob's particle is again in the state
\( wP[|0\rangle _{B}]+(1-w)P[|1\rangle _{B} \). Thus we conclude that any set
of commuting states can be teleported via a classically correlated channel.

Here one may note that this teleportation protocol is essentially a \( 1\rightarrow 3 \)
broadcasting protocol \cite{13} where the third particle is at a distant location,
and where there must be a further operation with \( \sigma _{x} \) on \( A_{2} \)
in the case when \( P_{2} \) clicks. This can be easily generalized to a \( 1\rightarrow N \)
broadcasting protocol by using the state \( \frac{1}{2}P[|000....0\rangle _{B_{1}B_{2}...B_{N-1}}]+\frac{1}{2}P[|111.....1\rangle _{B_{1}B_{2}...B_{N-1}}] \)
as the blank state where only two particles, the particle whose state is to
be teleported and \( B_{1} \) are required to be at the same location.

One can now relate the properties of the teleportation \emph{channel state}
(leaving aside the accompanying LOCC that are also required in any teleportation
protocol) with the \emph{set of states to be teleported} in the following way:\cite{14}

\noindent (1) For a set of orthogonal states (where cloning is possible), no
correlation (quantum or classical) is required in the channel state.

\noindent (2) For a set of commuting states (where no-cloning holds but broadcasting
is possible), classical correlation in the channel is sufficient for teleportation.

\noindent (3) For a set of noncommuting states (where even broadcasting is not
possible), an entangled channel state is necessary for teleportation.

Lastly we show that exact disentanglement of a set of \( 2\otimes 2 \) states
is possible when the reduced density matrices on one side are commuting \cite{16},
without applying the partial transpose operation \cite{17}. This easily follows
from the fact that the state of the side in which the density matrices commute,
can be teleported exactly through a separable channel. This fact along with
the above lemma implies that so far as local operations are concerned, any set
of \( 2\otimes 2 \) density matrices between two particles 1 and 2 can be exactly
disentangled \emph{if and only if} the reduced density matrices of at least
one party commute.

Recently some attempts has been made to understand whether quantum teleportation
is essentially a nonlocal phenomenon \cite{3,18}. In particular, Hardy \cite{3}
has shown that in general, teleportation is conceptually independent of nonlocality.
To show this he constructed a \textit{toy} model which is local (in the sence
that this model has a local hidden variable description) and in which no-cloning
holds, but still teleportation is possible. Then the question arises whether
there exists a bipartite state in quantum theory which has a local hidden variable
description but is still useful (as the channel state) for exactly teleporting
a set of states which cannot be cloned. In this letter we have shown that this
type of scenario really exists in quantum theory. There is a set of states (any
set of commuting states which is \emph{not} simply a set of orthogonal states)
which cannot be cloned but can be teleported through a classically correlated
channel state, which obviously has a local hidden variable description. We have
also shown here (in a protocol-independent way) that in exact teleportation
of any set of noncommuting states (where no-broadcasting \cite{13} holds in
addition to no-cloning), entanglement of the channel is necessary.

Based on the results till found, one would perhaps be inclined to think that
it is the no-cloning theorem in quantum mechanics which necessitates the use
of entanglement in teleportation channel. But we see here that it is not true.
Rather one can see that noncommutativity plays the fundamental role in deciding
the necessity of entanglement of the channel. This interplay between noncommutativity
and entanglement in teleportation can further be exploited to probe the more
difficult question as to whether quantum teleportation is a fundamentally nonlocal
phenomenon.

The authors acknowledge Debasis Sarkar for useful discussions. U.S. acknowledges
partial support by the Council of Scientific and Industrial Research, Government
of India, New Delhi.


\begin{thebibliography}{10}
\bibitem{1}C. H. Bennett, G. Brassard, C. Crepeau, R. Jozsa, A. Peres and W. K. Wootters,
\textit{Phys. Rev. Lett.} \textbf{70} (1993) 1895
\bibitem{2}M. B. Plenio and V. Vedral, \textit{Teleportation, Entanglement and Thermodynamics
in the quantum world}, quant-ph/9804075
\bibitem{3}L. Hardy, \textit{Disentangling Nonlocality and Teleportation}, quant-ph/9906123
\bibitem{4}L. Henderson, L. Hardy and V. Vedral, \textit{Two State Teleportation}, quant-ph/9910028
\bibitem{5}Of the qubits forming the two-qubit state \( \rho _{12} \), qubit \( 2 \)
was never operated upon and hence its state \( tr_{1}\left[ \rho _{12}\right]  \)
remains intact even after the teleportation protocol has been implemented. Qubit
\( 1 \) was exactly teleported to Bob and so its state \( tr_{2}\left[ \rho _{12}\right]  \)
must also remain the same after the teleportation process.
\bibitem{6}Here by No-Disentanglement theorem, we mean the No-Disentanglement into separable
states of \cite{7}. This is obviously stronger than the No-Disentanglement
into product states in \cite{8}.
\bibitem{7}T. Mor, \textit{Phys. Rev. Lett.} \textbf{83} (1999) 1451
\bibitem{8}D. Terno, \textit{Phys. Rev. A} \textbf{59} (1999) 3320
\bibitem{9}Recently Chau and Lo \cite{10} have shown (in a protocol-independent way) that
for universal exact teleportation, maximal entanglement of the channel state
is necessary, assuming that there is some \textit{a priori} entanglement in
the channel state. Therefore using our result, it follows that for universal
teleportation, maximal entanglement of the channel is necessary.
\bibitem{10}H. F. Chau and H. -K. Lo, \textit{How much does it cost to teleport?}, quant-ph/9605025
\bibitem{11}This has been proved in \cite{4}, but we include our independent proof as it
provides a clearer insight by using a new result on disentanglement of states
(which we prove in this letter).
\bibitem{12}Consider a randomly chosen state \( \rho _{AB} \) of two two-level systems
A and B from a given set S. If by any process (allowed by quantum mechanics),
the state \( \rho _{AB} \) turns into a separable state \( \rho ^{\prime }_{CD} \),
which is possessed by two two-level systems C and D (where (A,B) may or may
not be equal to (C,D)), such that \( tr_{B}\left[ \rho _{AB}\right] =tr_{D}\left[ \rho ^{\prime }_{CD}\right]  \)
and \( tr_{A}\left[ \rho _{AB}\right] =tr_{C}\left[ \rho ^{\prime }_{CD}\right]  \),
we then say that the set S can be disentangled.
\bibitem{13}H. Barnum, C. Caves, C. A. Fuchs, R. Jozsa and B. Schumacher, \textit{Phys.
Rev. Lett.} \textbf{76} (1996) 2818
\bibitem{14}By cloning, we mean transforming \( \rho  \) (along with a blank state and
a machine state) to \( \rho \otimes \rho  \) (after tracing out the machine).
This is possible if and only if \( \rho  \) is known to belong to a set of
orthogonal states \cite{15}. By broadcasting, we mean transforming \( \rho  \)
(of particle \( 1 \), say) (along with a blank state (of particle \( B \),
say) and a machine state) to a state \( \rho _{1B}^{\prime } \) (after tracing
out the machine) such that \( tr_{B}\left[ \rho _{1B}^{\prime }\right] =tr_{1}\left[ \rho _{1B}^{\prime }\right] =\rho  \).
This is possible if and only if \( \rho  \) is known to belong to a set of
commuting states \cite{13}.
\bibitem{15}W. K. Wootters and W. H. Zurek, \emph{Nature} \textbf{229} (1982) 802; D. Dieks,
\emph{Phys. Lett.} \textbf{92}\textbf{\emph{A}} (1982) 271; H. P. Yuen, \emph{Phys.
Lett.} \textbf{113}\textbf{\emph{A}} (1986) 405
\bibitem{16}T. Mor and D. Terno, \textit{Phys. Rev. A} \textbf{60} (1999) 4341
\bibitem{17}A. Peres, \emph{Phys. Rev. Lett.} \textbf{77} (1996) 1413; M. Horodecki, P.
Horodecki and R. Horodecki, \emph{Phys. Lett. A} \textbf{223} (1996) 1
\bibitem{18}S. Popescu, \textit{Phys. Rev. Lett.} \textbf{72} (1994) 792\end{thebibliography}
\end{document}